# Loss of New Ideas: Potentially Long-lasting Effects of the Pandemic on Scientists

Jian Gao[1,2,3], Yian Yin[1,3,4], Kyle R. Myers[5,6], Karim R. Lakhani[5,6,7], Dashun Wang[1,2,3,4*]

[1] Center for Science of Science and Innovation, Northwestern University, Evanston, IL 60208, USA
[2] Kellogg School of Management, Northwestern University, Evanston, IL 60208, USA
[3] Northwestern Institute on Complex Systems, Northwestern University, Evanston, IL 60208, USA
[4] McCormick School of Engineering, Northwestern University, Evanston, IL 60208, USA
[5] Harvard Business School, Harvard University, Boston, MA 02163, USA
[6] Laboratory for Innovation Science at Harvard, Harvard University, Boston, MA 02134, USA
[7] Institute for Quantitative Social Science, Harvard University, Boston, MA 02138, USA
[*] Correspondence to: dashun.wang@northwestern.edu

**Extensive research has documented the immediate impacts of the COVID-19 pandemic on scientists, yet it remains unclear if and how such impacts have shifted over time. Here we compare results from two surveys of principal investigators, conducted between April 2020 and January 2021, along with analyses of large-scale publication data. We find that there has been a clear sign of recovery in some regards, as scientists' time spent on their work has almost returned to pre-pandemic levels. However, the latest data also reveals a new dimension in which the pandemic is affecting the scientific workforce: the rate of initiating new research projects. Except for the small fraction of scientists who directly engaged in COVID-related research, most scientists started significantly fewer new research projects in 2020. This decline is most pronounced amongst the same demographic groups of scientists who reported the largest initial disruptions: female scientists and those with young children. Yet in sharp contrast to the earlier phase of the pandemic, when there were large disparities across scientific fields, this loss of new projects appears remarkably homogeneous across fields. Analyses of large-scale publication data reveal a global decline in the rate of new collaborations, especially in non-COVID-related preprints, which is consistent with the reported decline in new projects. Overall, these findings highlight that, while the end of the pandemic may appear in sight in some countries, its large and unequal impact on the scientific workforce may be enduring, which may have broad implications for inequality and the long-term vitality of science.**

The COVID-19 pandemic has disrupted the scientific enterprise [1-6]. In particular, researchers in the "bench" sciences, female scientists, and those with young children reported especially strong negative impacts on their productivity [5-9]. Now, more than a year into the pandemic, and with multiple vaccines developed, circumstances have evolved substantially. This raises an important question: has the pandemic's impact on scientists evolved as well?

To answer this question, we conducted a survey in January 2021, randomly sampling US- and Europe-based scientists across a wide range of scientific fields. Importantly, we adopted the same sampling strategy as a previous survey we conducted in April 2020 [5], which allows us to directly compare the results of the surveys at these two very different stages of the pandemic (SI S1). In the January 2021 survey, we asked scientists the same information we solicited in the April 2020 survey, including professional and demographic features. We also added new questions that compare their overall research activity and output in 2020 with 2019. The newly added questions include the number of new research publications, new submissions, new collaborators, and new research projects they started each year. In addition, we asked whether scientists worked on COVID-19 related research in 2020. In total, we collected responses from 6,982 respondents who self-identified as faculty or principal investigators (SI S2). To supplement our survey findings, we also conduct analyses using large-scale publication datasets, as recorded in the Dimensions database, which captures both articles and preprints published up to the beginning of 2021.

**The pandemic's impact remains significant but has shifted substantially in nature**

During the early phase of the pandemic, scientists reported a sharp decline in time spent on research [5, 6, 9]. For example, in April 2020, scientists on average reported a large initial decrease of 7.1 hours per week in work time, compared to pre-pandemic levels (Fig. 1a, left). In January 2021, however, scientists reported only minor differences between their current and pre-pandemic work time (Fig. 1a, right). Total work hours are still lower than the pre-pandemic levels, but the difference is now only 2.2 hours per week on average. In percentage terms relative to pre-pandemic levels, the impact on total work hours has changed from -13.8% in April 2020 to -4.3% by January 2021 (SI Figure S1), showing a clear sign of recovery.

Besides work time, previous research also identified changes in research submissions and publications during the early phase of the pandemic [10-12], prompting us to examine these metrics in the January 2021 survey. We find that, on average, the self-reported number of new publications or new submissions in 2020 is only moderately lower than in 2019 (Fig. 1b, left and middle). These reported changes are consistent with measurements from publication databases (SI Figure S9). Overall, the time- and publication-based metrics offer an encouraging sign of recovery. The relatively modest declines in submission and publication rates that we observed is consistent with prior studies [12, 13], some of which even identified increases in these metrics (SI Figure S9). Yet, as we show next, these short-term metrics mask a crucial dimension in which the pandemic has been impacting scientists: the rate of new research projects initiated.

**Fewer new projects initiated during the pandemic**

In terms of the pre-pandemic rate of new projects, only about 8.9% scientists reported that they initiated no new research project in the year of 2019, but this fraction has increased by roughly threefold, reaching about 27.0% in 2020 (Fig. 1b, right). Figure 1c plots the distributions of individual-level changes in the number of new publications, new submissions, and new projects for all respondents of the January 2021 survey. We find that the changes in new publications (Fig. 1c, left) and new submissions (Fig. 1c, middle) are rather modest, compared with the large, negative change in new projects; its distribution shows a clear shift to the left (Fig. 1c, right). These results are robust when measuring changes using absolute or logged values (SI Figures S2-S3).

Roughly one-third of survey respondents reported working on COVID-related research in 2020 (SI S9), echoing science's strong response to the novel pandemic [14, 15]. Yet, to the extent these were "new" projects for these scientists, it suggests that the majority of scientists who did not pursue COVID-related research may have experienced even greater declines. To investigate this possibility, we separate our samples based on whether or not scientists reported that they worked on COVID-related research in 2020. We find that the two groups show dramatically different patterns (Fig. 1d). For scientists who worked on COVID-related research, their responses across all dimensions that we surveyed—total work time, new publications, new submissions, or new projects—show almost no changes compared with the pre-pandemic level. Conversely, the "non-COVID" scientists reported clear decreases in total work time (-5.1%), new publications (-9.0%), and new submissions (-14.8%). Among these measures, the relative change in new projects shows the largest decline, averaging -36.2%. In absolute terms, this roughly corresponds to the loss of one new project per scientist in 2020 alone (SI Figure S2). Considering that scientists in our sample on average initiate only about three new projects in a normal year, this decline seems rather meaningful. When we examine these measures across the various stages of scientific production, shifting from a focus on finished papers to starting new research projects, it appears that the impact of the pandemic on scientific productivity magnifies as we move deeper into researchers' pipelines (Fig. 1d). In other words, there may be larger impacts of the pandemic that have not yet manifested in publication-based or other short-term metrics. These observations thus suggest that, as productive as science may appear at the moment, the pandemic's long-lasting effects on scientific productivity loom large on the horizon.

**Field- and group-level differences**

How does the decline in starting new research projects vary across various professional and demographic characteristics? To answer this question, we employ a Lasso regression approach to select features that are most predictive of changes in new projects (SI S3). First, we examine demographic features, after controlling for scientific fields and a COVID dummy variable capturing whether the respondent reported to engage COVID-related research. We find that the decline in new projects is sharply split across certain demographic dimensions. Specifically, the most important features associated with the declines in new projects are being a female and having young children (Fig. 1e). Notably, these are exactly the same groups of scientists who reported the largest initial disruptions to their research in the early phase of the pandemic [5, 6, 8], suggesting the loss of new projects may further exacerbate the pandemic's already highly unequal effects on scientists, especially for those who didn't pursue COVID-related research in 2020.

We next focus on differences in the rate of starting new projects across scientific fields. We again employ a Lasso regression model, but this time to select important fields while controlling for demographic features and the COVID dummy. We find that the decline in new projects affects research disciplines in a remarkably homogeneous way (Fig. 1f). Indeed, although research disciplines all reported substantial declines in starting new projects, especially for projects that are unrelated to COVID, only the biochemists reported significantly lower-than-average declines (post-Lasso regression coefficient $b = -0.12$, $S.E. = 0.05$, $p$-value $< 0.05$), after controlling for other individual-level features. Notably, this field-level homogeneity sharply contrasts the early impact of the pandemic on scientists, which was characterized by a high degree of heterogeneity across fields [5]. Indeed, in terms of the initial disruptions in work time, "bench" sciences or those requiring travel or in-person interactions are much more affected than "paper-pencil" disciplines [4, 5]. Given that the pandemic has halted access to lab facilities or travels to field sites, the level of homogeneity observed here is rather unexpected. Indeed, the January 2021 survey reveals that, despite the apparently different nature of work across fields, no scientific fields were immune to the loss of new ideas. These results are robust under several alternative measures (SI S7).

**An early sign for the loss of new projects: Decreases in new collaborations**

Given the long gestation time for new research ideas to mature and be published [13, 16, 17], this decline in new research projects suggests the true impact of the pandemic may not surface for years. Nonetheless, these findings prompted us to ask if there might be any telltale signals detectable in the publication records. One possible factor for the loss of new projects may be a reduction in the rate of new collaborations. Indeed, the pandemic and associated social distancing measures have halted in-person or spontaneous interactions in science to an extreme degree, drastically reducing chance encounters or other social interactions that might facilitate the flow of new research ideas and collaborations [18-20]. It thus raises a question of whether the rate of new collaborations among scientists might have decreased in 2020. If so, given the clear difference in the rate of new projects we observe between COVID and non-COVID scientists (Fig. 1d), one may also expect this effect to be substantially different between the two groups.

We test these hypotheses directly by analyzing large-scale publication datasets, including about 9.49 million articles and preprints published in 2019 and 2020. Specifically, we examine the rate of new collaborations for both COVID and non-COVID-related papers by calculating the fraction of new author pairs [21, 22] (SI S5). For both articles and preprints, we find that the fraction of new collaborations on COVID papers increased in 2020 (Fig. 2a), featuring a relative increase of roughly 40% above the 2019 level (Fig. 2bc), which are largely consistent with prior studies [23-25]. Yet at the same time, new collaborations on non-COVID papers exhibit markedly different patterns (Fig. 2a), showing a substantial decrease of about 5% below the 2019 level (Fig. 2bc). These measurements from data are broadly consistent with self-reported changes in new collaborations in our survey (SI Figure S8).

The decrease in the rate of new collaborations for non-COVID papers published in 2020 may seem unexpected, given that many of these collaborations may have started before the pandemic, suggesting the effect may be even stronger for collaborations that started later in time. To test this hypothesis, we focus on non-COVID preprints published in 2020. Given the publication delays, one might expect the decrease in new collaborations is more pronounced in preprints than in articles (SI Figure S10), and the effect might grow stronger over the course of 2020 as the pandemic unfolded. To test these predictions, we plot the temporal changes in the rate of new collaborations for non-COVID preprints published in 2020 relative to those published in 2019 (Fig. 2d). We find that the decline in new collaborations is more evident for those published in the second half of the year than the first half, suggesting the effect is especially pronounced for projects finished later in the year. Note that the observed patterns in new collaborations may flow from other social and institutional factors. Nonetheless, we find that the rate of new collaborations and new projects are strongly correlated with each other even after controlling for other factors (SI Tables S1-S2). Overall, these results provide supporting evidence consistent with the hypothesis that the rate of formalizing new ideas and starting new projects on non-COVID research has declined during the pandemic.

**Concluding remarks and discussions**

Taken together, our surveys reveal two important but contrasting perspectives. On the one hand, the trends of recovery from the initial impacts of the pandemic offer encouraging news, showing that scientists' initially large declines in work time have largely recovered as of early 2021, and the current decreases in paper submissions and publications appear relatively minor. On the other hand, however, our analyses suggest a novel dimension of the pandemic's impact that has thus far received little attention: even as scientists return to work, they have been substantially less likely to pursue new research projects, suggesting that the impact of the pandemic on scientists may last longer than is commonly imagined.

These findings are important for several reasons. While many researchers have focused their attention on scientists who pivoted their research to help understand the coronavirus and the pandemic [26], it is important to recognize that the majority of scientists did not pursue COVID-related research (SI S9), either by choice or necessity, and this group of scientists seem especially hard hit across various output measures. Although submissions and publications appear to be holding steady, if not on the rise [12, 13], these trends may reflect scientists exploiting old topics, writing up existing research, submitting drafts earlier than they would have otherwise [4, 27], writing more grant proposals than typical [28], or revisiting old data and reviving legacy projects that they would otherwise have not prioritized. While all of these could contribute to increases in short-term output measures, they suggest that counting publication trends alone may paint an incomplete, if misleading, picture of the productivity of the research enterprise.

While the decline in new research projects coincides with the decrease in new collaborations, many other factors may also play a role. For example, the pandemic has reduced access to research facilities or field sites to a drastic degree, which is essential for generating new hypotheses and data. The lack of in-person interactions that have historically facilitated mentorship and hands-on trainings may disproportionately affect junior researchers, who are important driving force for new ideas in science [29]. The evolving nature of the pandemic and associated intermittent lockdowns may deter scientists from pursuing projects that are more prone to interruptions.

The homogeneous nature of the decline in new projects across almost all fields suggests that the reasons for this particular effect may be due to disruptions that are common across fields, which include the lack of in-person interactions and the psychological toll caused by the pandemic [30, 31]. Although virtual conferences were held and digital watercoolers were created, in-person communication between colleagues was virtually non-existent across the disciplines. And to the second point, the rate of scientists reporting feeling fatigued or angry has doubled in 2020 [32]. Thus, our results are in line with existing research that shows the importance of face-to-face interactions for spurring new research projects [18-20, 33] as well as the negative impacts of crises on an individual's creativity [34]. Notably, many of these factors, including the lack of in-person interactions and the stress and anxiety induced by the pandemic, are not unique to the scientific workforce, suggesting that the patterns we observe here may also apply to other creative professions [35, 36].

Overall, these findings have important implications for science policy in the wake of the pandemic and beyond. First, they support the importance of face-to-face social interactions and collaborations [18], reinforcing the value of resuming in-person activities [37, 38]. While there could be substantial gains from certain aspects of science shifting online (e.g., virtual seminars reducing travel demands and bridging geographical gaps) [38, 39], and more studies are needed to understand tradeoffs of virtual and in-person events, there could be important social functions related to the formation of new ideas that virtual events do not facilitate. Furthermore, these results contribute to policy discussions, such as institutional bridge funds [39], to encourage social interactions, facilitate new collaborations [18], and promote new ideas. And the decrease in new collaborations raises further questions of whether the character of work may have shifted, given that fresh teams are especially important for producing novel research [21]. Ultimately though, successful rebuilding of the global research enterprise would also depend on how well policy makers and institutional leaders address and manage the mental-health challenges facing scientists [31, 32].

The decline in initiating new projects is particularly pronounced for women and caregivers of young children, which is consistent with related work [6, 8, 40]. Likely in response to these sorts of patterns, many institutional leaders implemented policies such as tenure clock extensions [39]. As institutions begin their phased return, it may be tempting for decision makers to evaluate short-term metrics to gauge research outputs and inform their subsequent policies. Yet our results suggest that these short-term metrics may mask long-lasting effects of the pandemic. Ignoring long-run consequences may have profound implications not just for the inequality of science [41-43] but also its long-term vitality.

While surveys can be an effective way to gather timely data during evolving circumstances, and the analyses of large-scale data provide supporting evidence, our analyses have several limitations. (1) Our two surveys cover only the European- and US-based institutions, which limits the geographic coverage of our analysis. Given the global disparity in managing the pandemic [15, 44, 45], it is imperative to expand our analyses to different countries and regions in future work. (2) Although the survey outcomes and actual research outputs show a high degree of correlation (SI Figure S4), there may be cognitive biases in the self-reporting of productivity metrics. (3) Compared with other measures of research productivity, such as work time, submissions, and publications, the number of new projects is a relatively novel metric, whose meaning may be interpreted differently by scientists from different disciplines. As such, our results call for future work to further investigate this novel dimension of productivity, which could help to enrich our understanding into the early phase of the research pipeline. (4) Our surveys do not capture health related information. For example, scientists may have different exposures to the virus, or have family members who are affected by the virus, or both, which may further distract them from research activities; yet these dimensions are not captured in our analyses. (5) The effects discussed in this paper are based on correlational evidence, leaving open questions about what exactly may be driving the decline in new projects that we observe.

In early 2021, the U.S. National Institutes of Health unveiled its $1 Billion "Long COVID" initiative [46] to understand the prolonged health consequences of infection. This vividly illustrates the importance of long-term thinking in our combat against COVID-19 challenges as their impacts may be profound and long-lasting. Likewise, it is vital for science funders and institutional leaders to pay similar attention to the long-term effects of the pandemic on the scientific enterprise—even when science appears to recover from its initial disruptions.

**Acknowledgements**

We thank Wei Yang Tham, Nina P. Cohodes, Peter E. Schiffer, and all members of the Center for Science of Science and Innovation (CSSI) at Northwestern University for invaluable help. This work is supported by the Air Force Office of Scientific Research under award number FA9550-19-1-0354, National Science Foundation SBE 1829344, the Alfred P. Sloan Foundation G-2019-12485 and G-2020-13873, the Peter G. Peterson Foundation Pandemic Response Policy Research Fund, and the Harvard Business School Division of Faculty Research and Development.

**Author contributions**

D.W., K.R.M. and K.R.L. jointly conceived the project. J.G., Y.Y., K.R.M. and D.W. designed the experiments. J.G., Y.Y. and K.R.M. collected data. J.G. and Y.Y. performed empirical analyses. All authors collaboratively interpreted results. J.G. and D.W. wrote the manuscript. All authors edited the manuscript.

**Competing interests**

The authors declare no competing interests.

**Human research participants**

The study protocol was approved by the Institutional Review Board (IRB) from Northwestern University and Harvard University. Informed consent was obtained from participants.

**Data availability**

Because of the sensitive nature of some variables collected by the surveys, the IRB-approved protocol does not permit individual-level data to be made unrestricted and publicly available. Researchers interested in obtaining restricted, anonymized versions of this individual-level data should contact the authors to inquire about obtaining an IRB-approved institutional data sharing agreement. This work also uses data sourced from Web of Science and Dimensions.ai. Researchers who wish to access raw data should contact the data sources directly.

**Code availability**

Code necessary to reproduce all plots and statistical analyses will be made freely available.

**Additional information**

Correspondence and requests for materials should be addressed to D.W.

# Figures

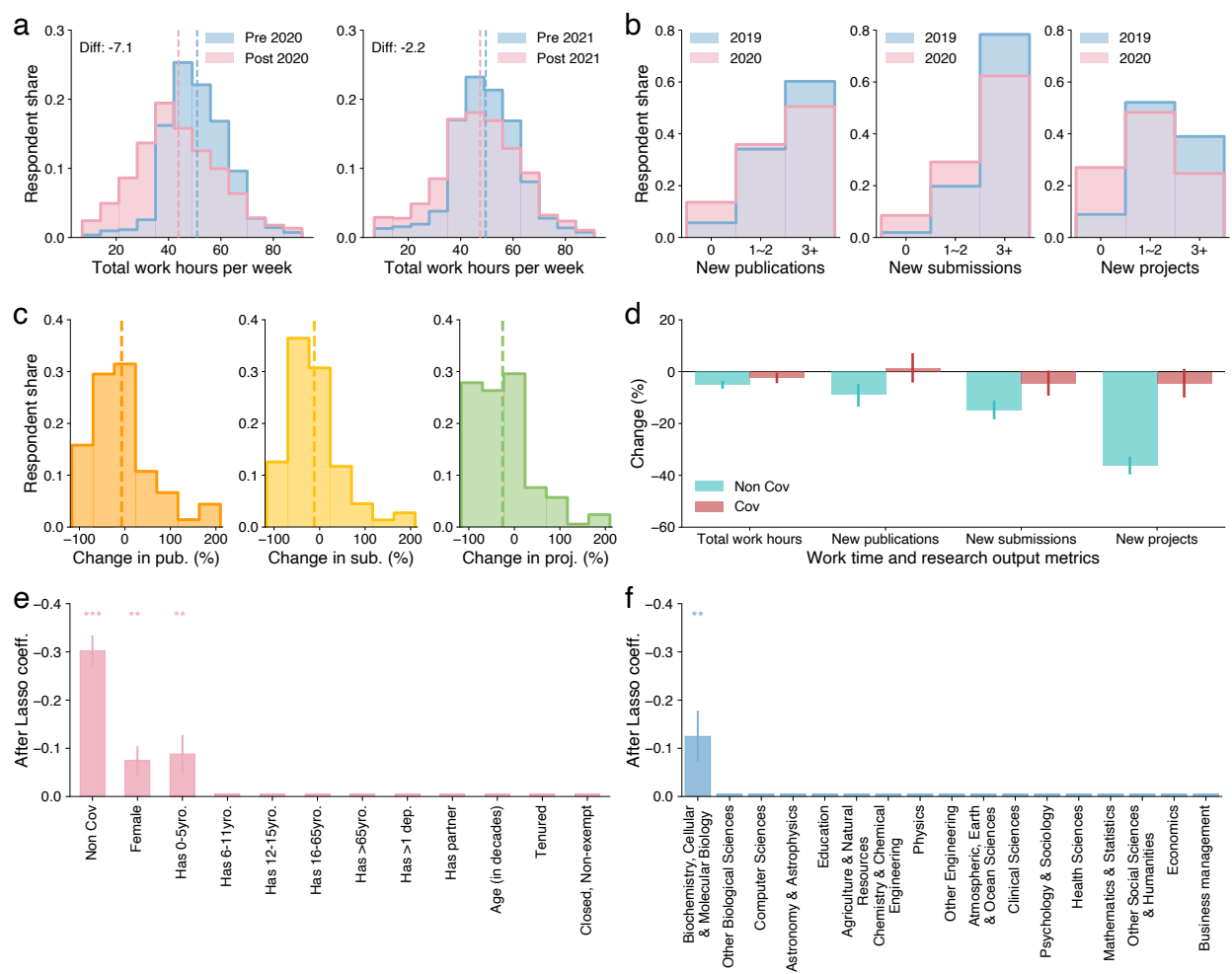

**Figure 1. Gradual recovery of total work time and substantial fewer new research projects.** (a) The distributions of total work hours per week for the pre and post periods. The left and right panels correspond to the surveys in April 2020 and January 2021, respectively. Vertical dashed lines mark the means, and the difference in means is shown. (b) The distributions of new publications, new submissions, and new projects for 2019 and 2020. Reported values are categorized into three bins. (c) The distributions of the changes in new publications, new submissions, and new projects in 2020 relatively to 2019. Changes over 200% are set as 200%. Vertical dashed lines mark the means. (d) The average change in work time and output metrics, unpacked by whether scientists have worked on COVID-related topics in 2020. Error bars indicate 95% confident intervals. (e) Regression analysis of the change in new research projects. After Lasso regression coefficients associated with important professional and demographic features selected by a Lasso approach after controlling for research fields. The regression also includes a COVID dummy variable capturing whether the respondent reported to engage COVID-related research in 2020. (f) After Lasso regression coefficients associated with important field features after controlling for demographic factors and the non-COVID dummy. Error bars indicate standard errors, and starts indicate significant levels: *p<0.1, **p<0.05, ***p<0.01.

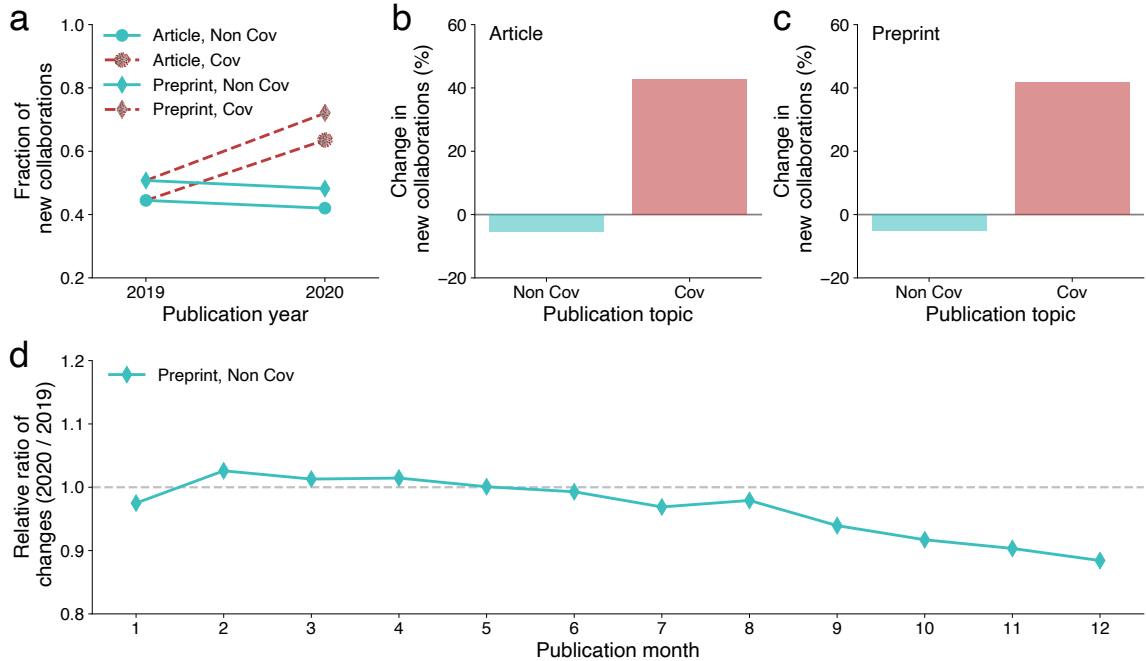

**Figure 2. Changes in new collaborations measured by large-scale publication datasets and survey responses. (a)** The fraction of new collaboration pairs in the author list of COVID and non-COVID papers published as articles or preprints in 2019 and 2020. **(b)** The average change in the fraction of new collaborations measured for articles in 2020 comparing with that in 2019. **(c)** The average change in the fraction of new collaborations measured for preprints in 2020 comparing with that in 2019. **(d)** The relative ratio of the fraction of new collaborations in 2020 over the fraction in 2019 measured for non-COVID preprints published in each month of the year.

# Supplementary Information

## Loss of New Ideas: Potentially Long-lasting Effects of the Pandemic on Scientists

Jian Gao[1,2,3], Yian Yin[1,3,4], Kyle R. Myers[5,6], Karim R. Lakhani[5,6,7], Dashun Wang[1,2,3,4*]

[1] Center for Science of Science and Innovation, Northwestern University, Evanston, IL 60208, USA
[2] Kellogg School of Management, Northwestern University, Evanston, IL 60208, USA
[3] Northwestern Institute on Complex Systems, Northwestern University, Evanston, IL 60208, USA
[4] McCormick School of Engineering, Northwestern University, Evanston, IL 60208, USA
[5] Harvard Business School, Harvard University, Boston, MA 02163, USA
[6] Laboratory for Innovation Science at Harvard, Harvard University, Boston, MA 02134, USA
[7] Institute for Quantitative Social Science, Harvard University, Boston, MA 02138, USA
[*] Correspondence to: dashun.wang@northwestern.edu

## Table of Contents

# S1. Survey Sampling and Recruitment

## 1.1 Web of Science corresponding authors

The Web of Science (WoS) publication database is leveraged to compile a large, plausibly random list of active scientists. We rely on the WoS database for two major reasons: (1) it is one of the most authoritative and widely used large-scale publication and citation corpuses available [1-5]; (2) it provides systematic coverage of corresponding author email addresses. Here, we attempt to focus on scientists that are likely to be in a more stable research position and still be active. Starting from about 21 million WoS papers published in the period of 2010-2019, we do an initial filtering based on the publication venue. Specifically, we exclude papers published in journals that are ranked to bottom 25% by the impact factor at the time (WoS Journal Citation Reports) for its WoS-designated category. We then extract all email addresses associated with these papers and consider an email address as a potential participant if (1) it is associated with at least two papers in the corpus, and (2) its affiliation of the most recent paper is based in the U.S. or Europe.

After this data filtering process, we are left with approximately 1.5 million unique email addresses, with about 521,000 in the U.S. and about 938,000 in Europe. We then randomly shuffled the two lists of email addresses separately and sampled 280,000 from the U.S. and 200,000 from Europe for the April 2020 survey [6]. As a part of a broader outreach strategy underlying this and other research projects, we oversampled the U.S. in comparison with Europe. Furthermore, from remaining email addresses in the two lists, we randomly sampled 140,000 from the U.S. and 100,000 from Europe for the January 2021 survey.

## 1.2 Participant recruitment

We recruited participants by sending invitations to the sampled email addresses. We followed the same recruitment process across the two surveys and used personalized texts to accommodate the time of the surveys. The recruitment email text used in the January 2021 survey is as follows:

> *Dear [Author Name],*
>
> *We need your help to shed light on how the coronavirus pandemic is affecting scientists like you. This study builds on our previous research published in Nature Human Behavior with Kyle R. Myers and Karim R. Lakhani at the Laboratory for Innovation Science at Harvard.*
>
> *Please take a brief moment to complete this short 5-minute survey as part of a research study. Your responses will help scientists and policymakers understand and respond to this rapidly evolving situation. The study protocol has been approved by Northwestern University's Institutional Review Board (IRB, STU00212699).*
>
> *Click HERE [hyperlink] to complete the survey or, copy and paste the URL below into your internet browser:*
> *[link]*
>
> *Thank you for your time!*
>
> *Sincerely,*
>
> *Dashun Wang, Ph.D.*
> *Northwestern Kellogg Center for Science of Science & Innovation (CSSI)*

# S2. Survey Instrument and Sampling Approach

## 2.1 Survey questions

The April 2020 survey includes questions on demographic information (age, gender, cohabitation, dependents), professional information (position type, institution type, fields of study, type of research, tenure status), and time allocation (time spent on different activities before and after the pandemic) [6]. In the January 2021 survey, besides retaining these questions, we added new questions about whether respondents worked on topics related to the coronavirus in 2020 and how their research output metrics changed in 2020 compared with 2019. Respondents were not required to answer any of the demographic questions. The January 2021 survey questions underlying the variables used in our analyses are as follows:

*Q. Which of the following best describes your current position?*
- *Faculty or principal investigator | Research staff or assistant | Post-doctoral researcher | Graduate student in a doctoral program | Retired faculty or principal investigator still engaged in research | Retired scientist no longer engaged in research | Other*

*Q. Which of the following best describes your field of study?*
- *[list of 20 fields]*

*Q. Which of the following best describes the institution you are primarily affiliated with?*
- *University or college | Non-profit research organization | Government or public agency | For-profit firm | Other*

*Q. Please answer the following:*
- *Is your institution physically closed to non-essential personnel?*
    - *Yes | No | Not relevant*
- *Are you exempt from the closure and allowed to travel to your work site(s)?*
    - *Yes | No | Not relevant*
- *Do you have tenure?*
    - *Yes | No | Not relevant*

*Q. Gender:*
- *Male | Female | Other | Prefer not to say*

*Q. Age:*
- *Under 20 | 20-24 | 25-29 ... 75-79 | 80 or older | Prefer not to say*

*Q. Number of dependents of any age you care for:*
- *0 | 1 | 2 | 3 or more | Prefer not to say*

*Q. In what age group(s) are your dependents? Note. You may select multiple*
- *0-2 years old | 3-5 years old | 6-11 years old | 12-18 years old | 19-65 years old | Over 65 years old*

*Q. Cohabitation status:*
- *I reside with a partner, spouse, or significant other | I reside with friends | I reside by myself | Other | Prefer not to say*

*Q. Around this time last year (January 2020), about how many hours per week did you work on anything related to your job? (e.g., researching, teaching, writing)*

- *14-21 hours per week (avg. 2-3 hours every day) | 21-28 hours per week (avg. 3-4 hours every day) |... | 77-84 hours per week (avg. 11-12 hours every day) | More than 84 hours per week (avg. 12 hours or more every day)*

*Q. Currently (January 2021), about how many hours per week are you working? (e.g., researching, teaching, writing)*

- *14-21 hours per week (avg. 2-3 hours every day) | 21-28 hours per week (avg. 3-4 hours every day) | ... | 77-84 hours per week (avg. 11-12 hours every day) | More than 84 hours per week (avg. 12 hours or more every day)*

*Q. During the year 2020, have you work on research topics related to this COVID-19 pandemic?*

- *No | Yes*

*Q. During the year 2019, how many new research projects did you start? Please write a number.*

- *[0, 1, 2, ..., 100]*

*Q. During the year 2020, how many new research projects did you start? Please write a number.*

- *[0, 1, 2, ..., 100]*

*Q. During the year 2019, with how many people did you establish new collaborations? Please write a number.*

- *[0, 1, 2, ..., 100]*

*Q. During the year 2020, with how many people did you establish new collaborations? Please write a number.*

- *[0, 1, 2, ..., 100]*

*For the following, please consider your "research publications" as all of your publications that focus on a research question. (e.g., journal articles, conference proceedings, patents, books. Ignore commentary, editorials, etc.)*

*Q. During the year 2019, how many research publications did you submit (including journals/conferences/preprint servers)? Please write a number.*

- *[0, 1, 2, ..., 100]*

*Q. During the year 2019, how many peer-reviewed research publications did you publish? Please write a number.*

- *[0, 1, 2, ..., 100]*

*Q. During the year 2020, how many research publications did you submit (including journals/conferences/preprint servers)? Please write a number.*

- *[0, 1, 2, ..., 100]*

*Q. During the year 2020, how many peer-reviewed research publications did you publish? Please write a number.*

- *[0, 1, 2, ..., 100]*

## 2.2 Research field definitions

Research fields in our survey are built on the field classifications used in national surveys such as the U.S. Survey of Doctorate Recipients (SDR) with an aggregation to ensure sufficient sample sizes within each field. We made additions to the fields by including Business Management, Education, Communication, and Clinical Sciences, as they reflect major schools at most universities and/or did not immediately map to some of the default fields used in the SDR, for example, medical specialties are not included in the "Health Sciences" field in SDR.

### 2.3 Survey data sampling approach and basic statistics

After a total of 480,000 emails sent in April 2020, there are 8447 individuals that entered the survey and continued past the consent stage. For our analysis, we focus entirely on responses from faculty or principal investigators (PIs). Thereby, we retain respondents who self-identified as "Faculty principal investigator" or "Retired faculty or principal investigator still engaged in research" and reported working for a "University or college", "Non-profit research organization", "Government or public agency", or "Other" (excluding whose who reported working for a "For-profit firm"). We further drop observations that have missing data for key variables (working time, age, gender, and field of study). We do not impute missing variables as it may introduce unnecessary noise [7]. Altogether, these criteria lead to a sample of 4535 respondents used in the analyses for the April 2020 survey.

Following a similar procedure, we sent out our January 2021 survey to 240,000 email addresses, and 4672 individuals entered the survey and continued past the consent stage. We focused on responses from faculty or principal investigators (PIs) and dropped observations that have missing data for working time. These criteria lead to a base sample of 2447 respondents from the January 2021 survey in our analysis. We further drop missing data for research output metrics (i.e., projects, collaborators, submissions, and publications) and key variables (e.g., age, gender, field of study, tenure status, etc.) during the analyses involving them.

## S3. Covariate Selection and Regression Approach

### 3.1 Lasso selection and post-Lasso regression

We use multivariate regressions to explore whether changes associated with a group of individuals change after conditioning on other observables (e.g., the demands of home life unique to certain individuals, or the nature of work in certain research fields). We select a set of important covariates (or transformations thereof) that should be included in regressions by employing a Lasso method, which provides a data-driven approach to this selection problem by excluding covariates from the regression that do not improve the fit of the model [8, 9]. Specifically, our Lasso approach is to include a vector of indicator variables for the research fields and the professional and demographic groups of interest. When focusing on field-level differences, we include the professional and demographic variables in the control set. In turn, when focusing on professional and demographic-level differences, we include field variables in the control set. To make minimal assumptions about the functional form of control variables, we conduct the following transformations to expand the set of controls: for all continuous variables we use inverse hyperbolic sine (which approximates a logarithmic transformation while allowing zeros), square and cubic transformations, and we interact all indicator variables with the linear versions of the continuous variables.

The Lasso selection approach is performed using the lasso linear package in Stata 16 © software. We use the defaults for constructing initial guesses, tuning parameters, number of folds (ten), and stopping criteria. We use the two-step cross-validation "adaptive" Lasso model where an initial instance of the algorithm is used to make a first selection of variables, and then a second instance occurs using only variables selected in the first instance. The variables selected after this second run are then used in a standard post-Lasso Ordinary Least Squares (OLS) regression with heteroskedastic robust standard errors.

### 3.2 Ordinary Least Squares (OLS) and probit regression

A standard OLS regression model is used to explore the relationship between two variables of interests, in our analysis, the association between the change in new projects and the change in new collaborators. We add a set of control variables in the OLS regressions with robust standard errors, including both professional and demographic variables and the dummy of research fields. Moreover, we employ a probit regression model to study the how these variables are associated with the probability for scientists to work on COVID-related research. The dependent variable is a dummy variable that takes 1 if scientists reported working on COVID-related research in 2020 and 0 if otherwise. The independent variables include professional and demographic variables as well as the field dummy.

## S4. Publication Data and Matching Approach

### 4.1 The Dimensions publication data

For the database of scientific publications, we use Dimensions [10], a data product by Digital Science, which provides a systematic coverage of research papers and preprints. The Dimensions data is updated in a timely manner with relatively smaller time lags compared with other alternative large-scale publication datasets, offering an opportunity to study recent publishing trends. In March 2021, we retrieved from the Dimensions database papers and preprints published up to the end of 2020. For each paper, we obtain information on its title, author list, publishing venue, publication date, fields of study, DOI (Digital Object Identifier), and publication type (e.g., articles and preprints). Authors in the corpus have been pre-disambiguated by Dimensions.

We also construct a set of publications that are related to the COVID-19 pandemic by leveraging the Dimensions searching engine. Specifically, we follow prior work [11] and search for papers published in 2020 using the following query suggested by Dimensions [12]:

> *"2019-nCoV" OR "COVID-19" OR "SARS-CoV-2" OR "HCoV-2019" OR "hcov" OR "NCOVID-19" OR "severe acute respiratory syndrome coronavirus 2" OR "severe acute respiratory syndrome corona virus 2" OR (("coronavirus" OR "corona virus") AND (Wuhan OR China OR novel))*

This searching process yields in total 216,187 COVID-related papers published in 2020 out of all papers indexed by the Dimensions database.

### 4.2 Matching survey respondents to Dimensions authors

We link the respondents of January 2021 survey to authors in Dimensions. For each respondent, we first build a list of papers that are associated with the respondent's email address in the WoS dataset and collect each paper's DOI, one of the most commonly used identifiers for scientific publications. Then, we retrieve papers from Dimensions using the list of DOIs and collect author information for each paper. Next, we aggregate the author information and identify the Dimension author that can be matched to the respondent on both first and last names. Using this matching process, we linked 2141 out of all 2447 respondents in the January 2021 survey to Dimensions authors, yielding a matching rate of about 87%. For these matched respondents, we further calculate the number of their publications every year in Dimensions.

## S5. Measuring the Rate of New Collaborations

### 5.1 Measuring new collaborations in publication data

To investigate the temporal changes in collaborations among scientists, we extract co-authorship patterns from the Dimensions publication data and calculate a measure of new scientific collaborations. Specifically, building on rich literature of team science [5, 13-17], we define the rate of new collaborations as the fraction of author pairs that have not collaborated previously to all authors pairs on a paper. Considering that the baseline rate of new collaborations may change as team size (i.e., the number of authors on a paper) increases, here we only focus on teams with 50 or less authors [18]. We repeat our analysis for narrow ranges of team sizes as additional robustness checks (Figure S11).

To construct a comprehensive record of previous collaborations, we extracted all disambiguated author pairs from publications in Dimensions since 1950. For each author pair, we record their earliest year of collaboration (i.e., the time of their first coauthored publication). Then, we iterate over all coauthor-paper combinations, classifying each coauthor pair as an "old" collaboration if their earliest collaboration year is before the publication year of this paper. For example, given a paper published in 2020 with three authors Alice, Barbara and Cindy, if Alice has published one paper with Barbara in 2019 and another paper with Cindy in 2018, but Barbara and Cindy haven't co-authored any paper before 2020, the fraction of new collaborations in this paper is 1/3. We also use information on publication month as well as publication types for our further analysis.

## S6. Changes in Time and Output Metrics

### 6.1 Changes in total work hours

We investigate the change in work time (comparing post-pandemic levels with pre-pandemic levels) and explore how the magnitude of changes shits over time as the pandemic unfolds in 2020. Figure S1 shows the changes in total work hours reported by respondents in April 2020 survey and January 2021 survey, which are 9 months apart. We find that the total work hours per week in the post pandemic period increases from an average of about 44 hours in April 2020 to about 47 hours in January 2021 (Figure S1a), largely narrowing the gap between pre and post pandemic work hours. The distribution of changes in total work hours at the individual level in April 2020 exhibits a clear shift to the left, with a mean decrease of about 7 hours. By comparison, the distribution in January 2021 becomes more symmetric, showing only a minor decrease of about 2 hours on average (Figure S1b). The percentage change in work hours has also increased from about -14% to about -4%, again suggesting a clear recovery pattern in working time (Figure S1c).

### 6.2 Changes in publication, submission, and projects

We calculate the changes in research output metrics reported by respondents in the January 2021 survey. Figure S2 shows the average values and the relative changes in the number of new research publications, new submissions, and new projects. We find that research outputs in 2020 are on average less than those in 2019 across all three metrics (Figure S2a). More specifically, the average number of new publications, submissions and projects has decreased by 0.4 (from 4.5 in 2019 to

4.1 in 2020), 0.9 (from about 6.2 in 2019 to 5.3 in 2020) and 0.7 (from about 2.5 in 2019 to 1.8 in 2020) approximately. In a relative term, the decline in new projects (-26%) is more pronounced than in new submissions (-5%) and publications (-11%) (Figure S2b). These results are largely robust when calculating changes using a logged value (Figure S2cd), where we use the form log(x+1) to deal with 0 in the raw value x.

For the calculation of relative changes (Figure S2cd), a small fraction (about 3-5%) of respondents reported a zero value for 2019 but a non-zero for 2020 are excluded from the analyses because the denominator can't be zero. To test the robustness of our results, we also calculate the absolute changes in the number of new publications, new submissions, and new projects comparing 2020 with 2019 (Figure S3). We find that all the three research output metrics show a negative change on average, providing additional support to our results.

### 6.3 Comparison of the publication data in survey and Dimensions

To test whether the self-reported number of publications is aligned with that tracked by other data sources, here we calculate publication counts for respondents matched in the Dimensions publication database (S4). For each of those with matched survey respondents, we calculate the number of articles published in 2019 and 2020 and compare with the number reported in the survey (Figure S4). We find that our survey data is strongly and positively correlated with the Dimensions data, showing a Spearman's rank correlation of 0.71 for 2019 (Figure S4a) and 0.75 for 2020 (Figure S4b). A linear fit of the two data can explain about 57% of variances, indicating an overall well alignment. Notably, the number of publications by Dimensions tends to be larger than the survey-reported number especially for more productive respondents, which may be explained by several reasons, such as recalling biases when respondents published a large number of papers and cognitive biases as respondents may have different perceptions of authorship when there are many coauthors on a single paper. As robustness checks, we also examine the relationship based on logged values, where we use log(x+1) to avoid 0 in the raw value x. We find that these positive correlations are very robust when using logged values (Figure S4cd).

## S7. Changes in Projects and Collaborators by Survey

### 7.1 Group-level and field-level changes in new projects

We explore the differences in the decline of new research projects in 2020 relative to 2019 across demographic variables and scientific fields. Figure S5 shows the average changes in new projects, aggregated by professional and demographic variables as well as fields. We find that scientist who haven't pursue COVID-related research reported a much larger declines in new projects, showing a change of about -57%, compared with the overall sample average of about -27% (Figure S5a). Female scientists and those with young dependents are also affected more than others. In addition, we find that scientists in disciplines that rely on physical laboratories—such as biochemistry, biology, and astronomy—reported the largest loss of new ideas in 2020, in a range of 38-44% below the 2019 level (Figure S5b). By comparison, scientists in social science fields (e.g., business, economics, and humanities) reported smaller declines in new ideas in 2020 compared with 2019.

These first-order observations may reflect a multitude of factors, including the composition of researchers in each field, the probability of working on COVID, etc. To address this issue, we also employ a Lasso regression approach to select important features that are most predictive of the changes in new projects with controlling for other factors. The regression results are reported in Fig. 2 of the main text and S8.2.

### 7.2 Robustness checks using alternative measures

In our main analyses, we use the percentage change in the number of new projects, comparing 2020 with 2019, as the primary measurement of changes in projects. Here we show some robustness checks on the results of group-level and field-level differences by using alternative calculation methods for the changes. First, we calculate the percentage changes in projects using logged values, where logged (x+1) is used to avoid 0 in the raw value x (i.e., the number of project). Figure S6 shows the average changes and the Lasso regression results based on the percentage changes in logged values. We find that the non-COVID dummy, female, and having young dependent are still the most important features that are predictive of declines in projects (Figure S6ab). The changes associated with research fields are relatively small comparing with the sample average, and only the biochemists reported significantly larger declines conditional on other factors (Figure S6cd). These observations are highly consistent with the results reported in Fig. 2 of the main text.

We also repeat our analyses using a measure of absolute changes in raw values from 2019 to 2020 (Figure S6e-h). We find that the results remain largely consistent, with only one additional demographic feature selected by the Lasso approach, i.e., having 6-11 years old dependents together with having 0-5 years old dependents (Figure S6f). Overall, these results support the robustness of our findings that the effects of losing new projects apply almost universally across fields but split sharply along some demographic dimensions.

### 7.3 Changes in new collaborators

We analyzed the changes in new collaborators based on the January 2021 survey response (Figure S7). We find that scientists reported a large decline in new collaborations in 2020. While only about 15% of scientists reported no new collaborators in 2019, but this fraction becomes more than twice as large at about 35% for 2020 (Figure S7a). On average, scientists reported about 3.9 new collaborators in 2019, while this number reduces to about 2.9 in 2020 (Figure S7b). The distribution of absolute changes in new collaborators is slightly left-shifted, with a mean and median value being around -1, suggesting a loss of one new collaborator on average in 2020. In a relative term, the decline in new collaborators is more striking, showing a percentage change of -17% on average. These observations are robust when we calculate changes in new collaborators based on logged values.

Moreover, we find that the rate of new collaborators differs massively, however, between COVID and non-COVID scientists (Figure S8). Specifically, COVID scientists reported a similar number of new collaborators in 2019 and 2020, while non-COVID scientists reported a substantial decline that the average number of new collaborators halves from about 4.4 in 2019 to about 2.2 in 2020 (Figure S8a). There is a clear left shift in the distribution of absolute changes in new collaborators for non-COVID scientists, with a large negative mean value of -1.5 (Figure S8b). By comparison, the distribution for COVID scientists is symmetrical, with a slightly positive mean value. In a relative term, non-COVID scientists reported about 32% reduction in new collaborators in 2020 (Figure S8c), while COVID scientists reported a net increase in new collaborators in 2020, with an average change of about 15% above the 2019 level (Figure S8d).

## 7.4 Associations between new projects and new collaborators

We use an OLS regression model to examine the associations between new projects and new collaborators. Table S1 summarizes the regression results, where we regress the number of new projects against the number of collaborators with controlling for professional and demographic variables and research fields. We find that the number of new collaborators has a significantly positive effects on the number of new projects even conditional on other important factors. We repeat this regression analysis by using different measures and find the results are robust. Specifically, the absolute changes in new projects are positively associated with the absolute changes in new collaborators, and the percentage changes in new projects are positively associated with the percentage changes in new collaborators (Table S2), showing consistent evidence that new projects and new collaborators are significantly and positively associated conditional on other important factors.

## S8. Results based on the Dimensions Publication Data

### 8.1 Trends of total and average publications

We analyze publication trends based on the large-scale publication data. Figure S9ab shows the total number of papers published in 2019 and 2020 as tracked by the Dimensions database. We find that the volume of publications increases in 2020 compared with 2019 for both articles (Figure S9a) and preprints (Figure S9b). This trend remains after excluding COVID-related publications in 2020, suggesting that scientists as a whole still published more non-COVID articles and preprints in 2020 than in 2019.

When considering the change in research output at the individual level, however, the data shows a different trend. Figure S9c shows the average number of articles in 2019 and 2020 separated by COVID authors (i.e., those who published COVID-related articles in 2020) and non-COVID authors (i.e., those who didn't published COVID-related articles in 2020). We find that COVID authors on average published much more articles in 2020 than in 2019, while for non-COVID authors the average number of articles in 2020 is very close to that in 2019. To align with survey respondents who are faculty or principal investigators, we further restrict the analyses for more active and senior authors who published at least one paper per year during the past five years (2015-2019). We find that COVID authors still published more articles in 2020, but non-COVID authors published slightly less articles in 2020 on average compared with 2019 (Figure S9d), which is consistent with our observations from the January 2021 survey that non-COVID scientists (i.e., those who worked on COVID-related research in 2020) reported a decline in the number of new publications in 2020 compared with 2019. Further, we vary the time window from five years to one year (Figure S9e) and ten years (Figure S9f), finding that the results are robust.

### 8.2 Changes in new collaborations measured by publication records

Using the Dimensions publication data, we further calculate the fraction of new collaborations for peer-reviewed articles and preprints published in the past decades (2010-2020). For the year 2020, we further separate the calculation by COVID-related papers and non-COVID papers (Figure S10a). We find that the curves are relatively flat from 2010 to 2019 for both articles and preprints. At the same time, there is a notably decrease from 2019 to 2020 for non-COVID papers and a substantial increase for COVID papers, showing a distinguish pattern in 2020.

We further calculate the relative ratio of the fractions of new collaborations for non-COVID preprints comparing the later year with the former year by month for two successive years (Figure S10b), e.g., dividing the fraction of 2020 by that of 2019. We find that the ratios for 2019/2018 and 2018/2017 remain largely stable across the year with a value being slightly larger than 1, indicating a minor yet stable increase in new collaborations over time. Notably, only the 2020/2019 ratio shows a clear decreasing trend during the second half of the year, suggesting a substantial decline in new collaboration on non-COVID preprints. Furthermore, considering the fraction of new collaborations can be affected by team size, we separate the analyses by the number of authors on a paper (Figure S11). We find that our observations are largely robust across different group sizes, and the effects tend to be more pronounced for small groups compared with large groups.

## S9. Survey Data on COVID Research

### 9.1 COVID-related research by survey data

A growing number of COVID-related publications indicate the strong response across the scientific community to the novel pandemic. According to the January 2021 survey, about 34% of survey respondents reported working on COVID-related research topics in 2020. We further explore the differences in worked on COVID across professional and demographic dimensions and research fields. Figure S12 shows the fraction of respondents that worked on COVID in 2020. We find that there are relatively small variations across professional and demographic dimensions compared with the sample average (Figure S12a). By comparison, there are notable differences across research fields, where a larger fraction of both social and clinical scientists but a small fraction of physicists worked on COVID (Figure S12b).

We further use a probit regression model, which regress worked on COVID in 2020 against professional and demographic variables and the research field dummy (Figure S13), finding our conclusions remain robust. Together, these results indicate that whether worked on COVID is largely associated with fields rather than with professional and demographic factors such as gender, parenthood, and tenure status.

# SI Figures

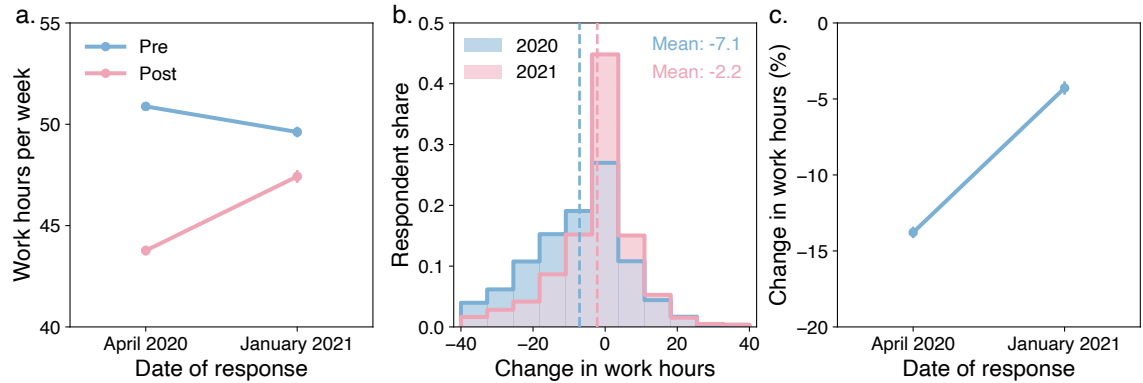

**Figure S1. Change in total work time.** (a) The total works hours per week in pre and post periods across two surveys in April 2020 and January 2021, respectively. (b) The change in total work hours per week comparing post period with pre period in April 2020 and January 2021 surveys. The dashed vertical lines mark the means. (c) The average percentage change in total work hours per week. Error bars indicate standard errors.

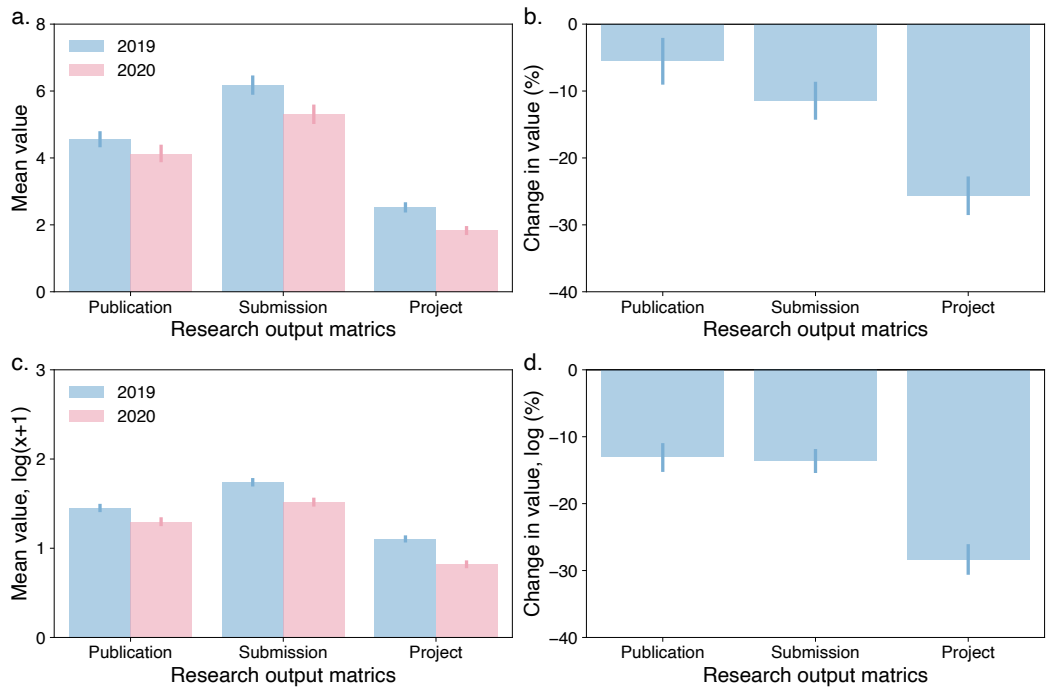

**Figure S2. Relative changes in research output metrics.** (a) The average number of new publications, new submissions, and new projects in 2019 and 2020, respectively. (b) The average percentage change in publications, submissions, and projects, comparing 2020 with 2019 at the individual level. (c-d) The results based on logged values. The raw value x is logged by log(x+1) to avoid 0. Error bars indicate 95% confident intervals.

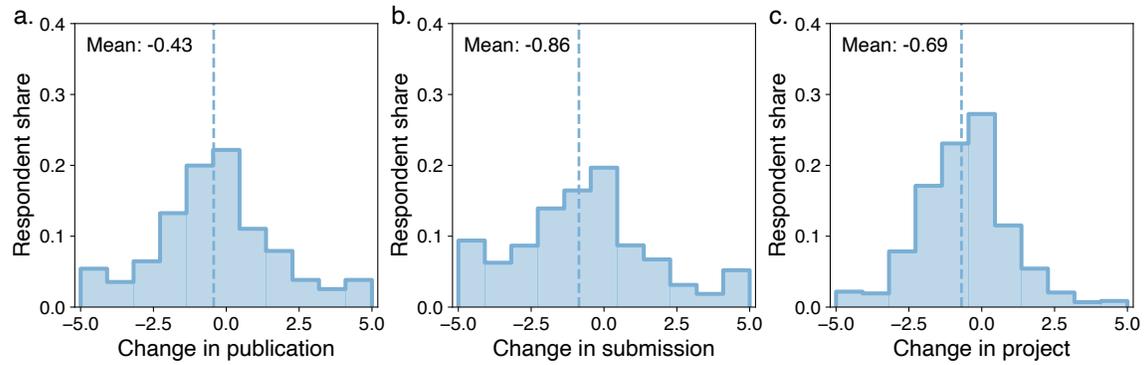

**Figure S3. Absolute changes in research output metrics.** The change in the number of (a) new publications, (b) new submissions, and (c) new projects, comparing 2020 with 2019. Changes that are below -5 or above 5 are set as -5 and 5, respectively. Vertical dashed line indicates the mean change.

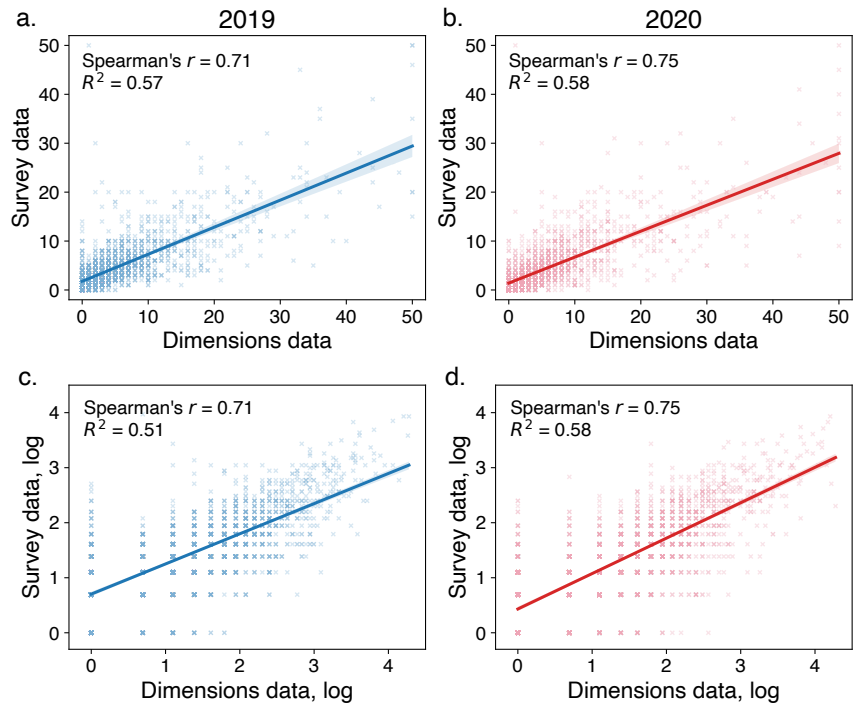

**Figure S4. Comparison on publications between the survey responses and the Dimensions data for matched scientists.** (a) The Spearman's correlation between the numbers of publications for 2019. (b) The Spearman's correlation between the numbers of publications for 2020. (c-d) show results based on logged values. Solid lines indicate linear fits with 95% confidence intervals. The goodness of fit is shown by R2.

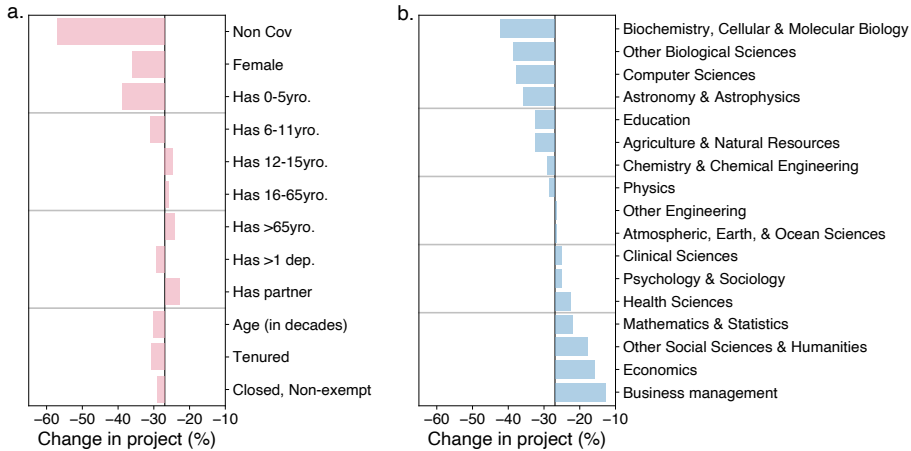

**Figure S5. Group-level and field-level changes in new projects comparing 2020 with 2019.** (a) The percentage changes in new projects, aggerated by professional and demographic dimensions. The non-COVID dummy takes 1 if scientists didn't work on COVID-related topics in 2020 and 0 if otherwise. (b) The percentage changes in new projects, aggerated by research fields. The changes are centered by the mean.

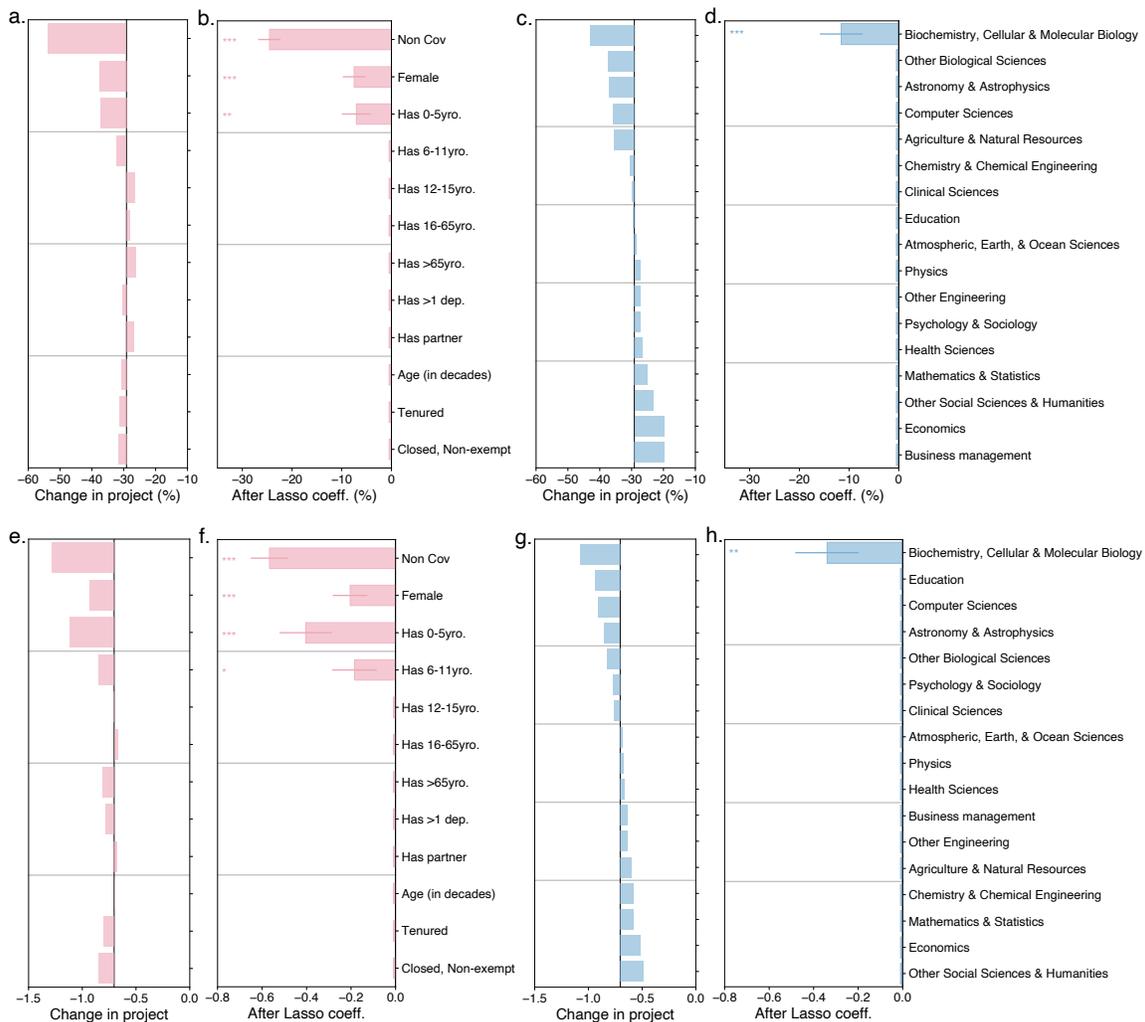

**Figure S6. Group-level and field-level changes in new projects based on logged and absolute values.** (a) The percentage changes in new projects that are aggerated by professional and demographic dimensions.

The changes are centered by the mean value. (b) After Lasso regression coefficients associated with important professional and demographic features selected by a Lasso approach after controlling for research fields. The regression also includes a non-COVID dummy variable that takes 1 if scientists didn't work on COVID-related topics in 2020 and 0 if otherwise. (c) The percentage changes in new projects that are aggerated by research fields. The changes are centered by the mean value. (d) After Lasso regression coefficients associated with important field features after controlling for demographic factors and the non-COVID dummy. (e-h) show the results based on the differences in raw values between 2020 and 2019. Error bars indicate standard errors, and starts indicate significant levels: *$p<0.1$, **$p<0.05$, ***$p<0.01$.

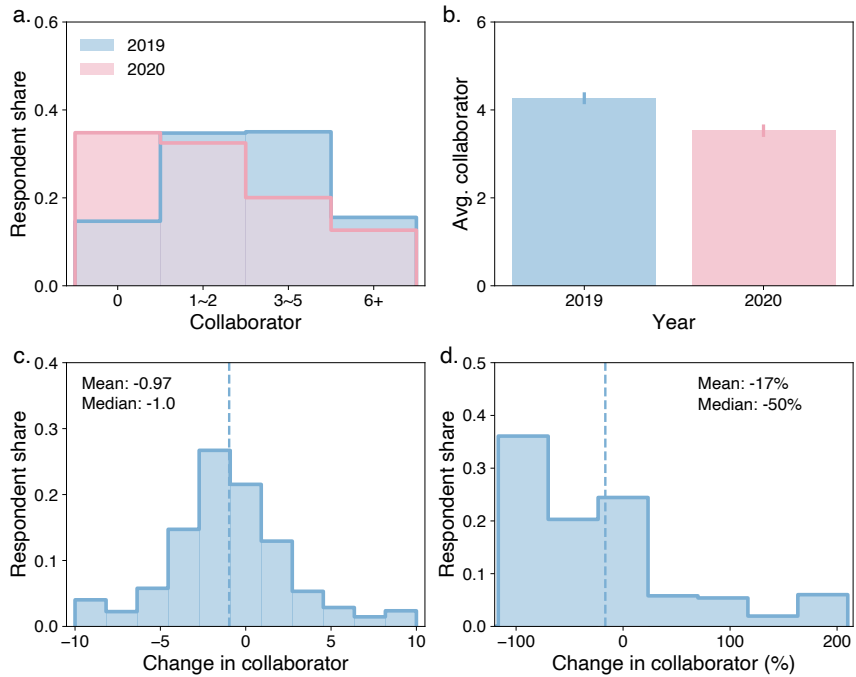

**Figure S7. Changes in new collaborations measured by survey responses.** (a) The distributions of new collaborators in 2019 and 2020 based on the January 2021 survey. (b) The average number of new collaborators in 2019 and 2020. (c) The absolute changes in new collaborators comparing 2020 with 2019. Changes that are below -10 and above 10 are set as -10 and 10, respectively. (d) The distributions of the percentage changes in new collaborators. Changes over 200% are set as 200%.

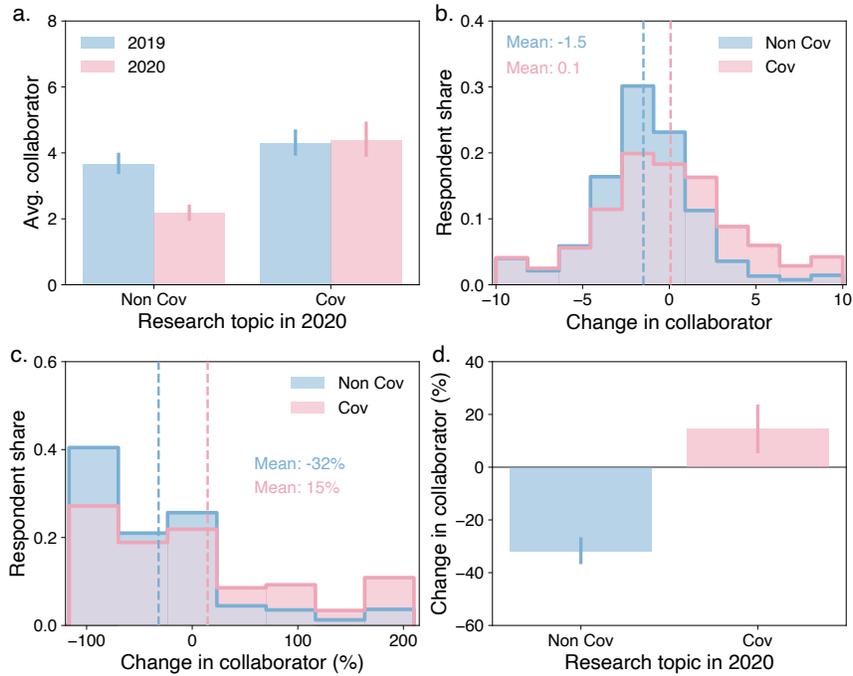

**Figure S8. Changes in new collaborations for COVID and non-COVID scientists measured by survey responses.** (a) The average number of new collaborators in 2019 and 2020. Error bars indicate 95% confidence intervals. (b) The absolute changes in new collaborators comparing 2020 with 2019. Changes that are below -10 and above 10 are set as -10 and 10, respectively. (c) The distributions of the percentage changes in new collaborators. Changes over 200% are set as 200%. Vertical dashed lines mark the means. (d) The average change in the number of new collaborators in 2020 compared with 2019 for COVID and non-COVID scientists. Error bars indicate 95% confidence intervals.

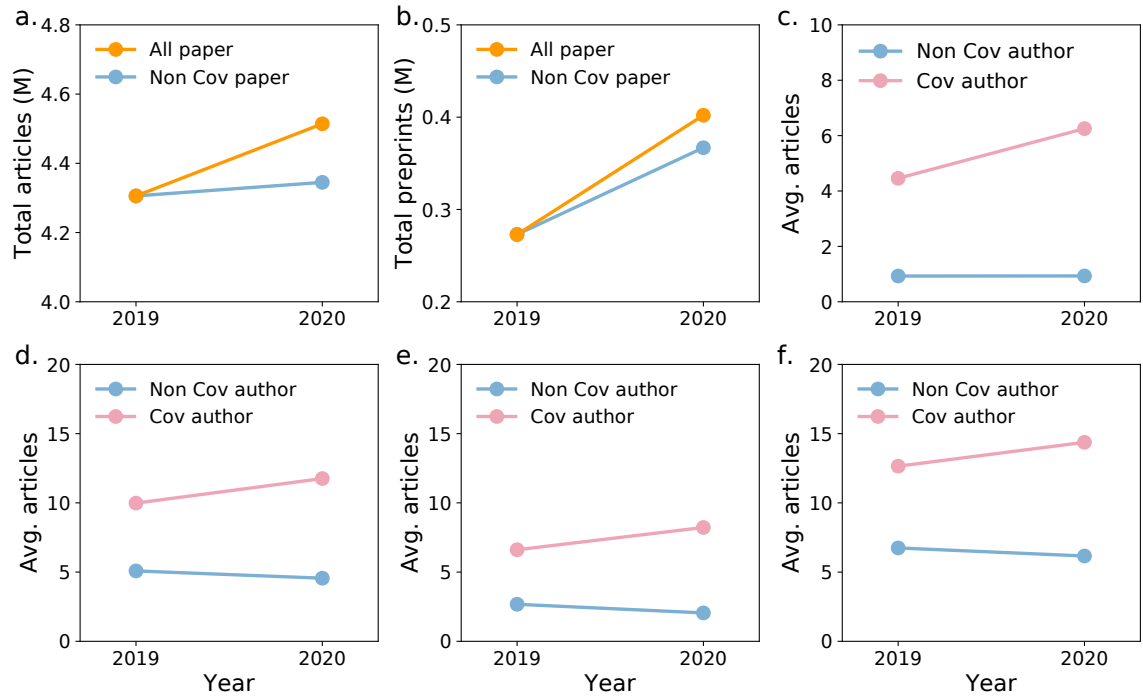

**Figure S9. The total and average number of publications in 2019 and 2020 according to Dimensions.** (a) The total number of all articles and non-COVID articles. The gap indicates the number of COVID-related articles published in 2020. (b) The total number of all preprints and non-COVID preprints. The gap indicates the number of COVID-related preprints published in 2020. (c) The average number of articles in 2019 and 2020 for COVID and non-COVID authors at the individual level for the full sample. (d) Results for authors that published at least one article per year during 2015-2019. (e) Results for authors that published at least one article in 2019. (f) Results for authors that published at least one article per year during 2010-2019.

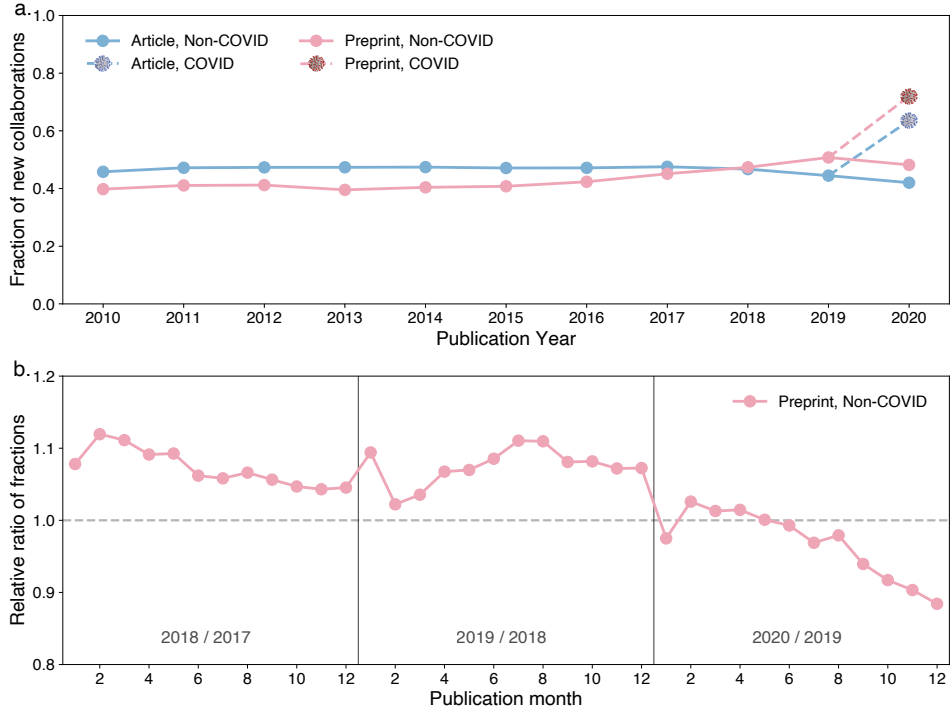

**Figure S10. The changes in new collaborations based on publications.** (a) The temporal changes in the fraction of new collaborations for papers published in a period of 2010-2020. The results are grouped by peer-reviewed articles (in blue) and preprints (in pink). The results are further separated by whether authors published COVID-related papers in 2020, namely, non-COVID authors (solid line) and COVID authors (dashed line). (b) The relative ratio of the fractions of new collaborations in non-COVID preprints published in two successive years. The results show three comparisons in the rate of new collaborations, namely, 2018 vs 2017, 2019 vs 2018, and 2020 vs 2019.

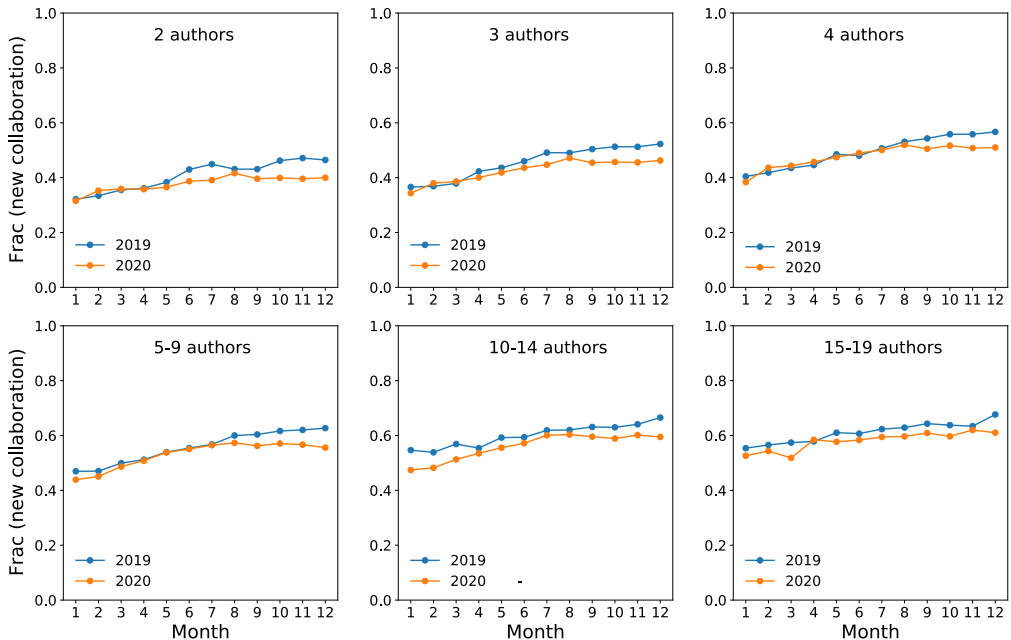

**Figure S11. The fraction of new collaborations measured by month for preprints published in 2019 and 2020.** The panels show the results for different group sizes, i.e., the number of authors on a preprint.

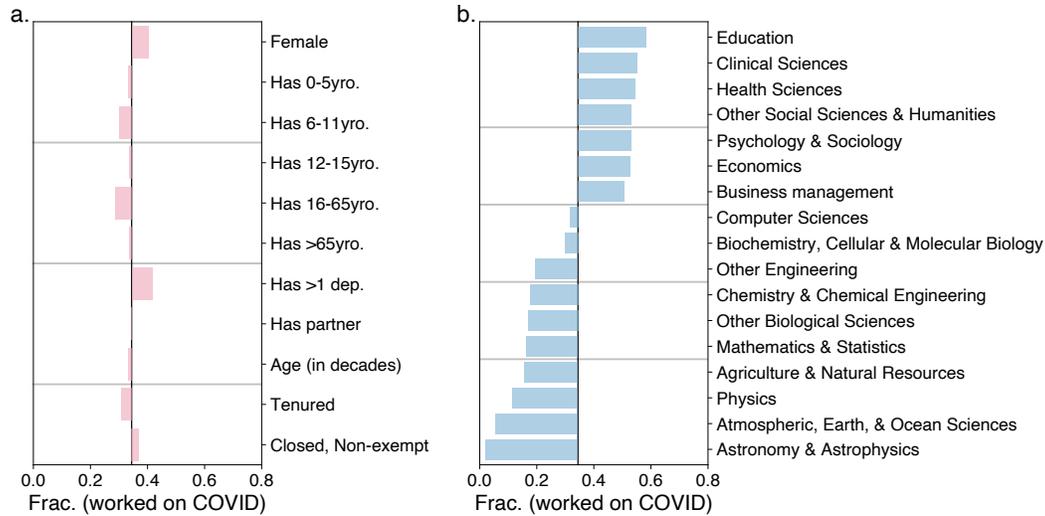

**Figure S12. The fraction of respondents that worked on COVID-related research in 2020.** (a) Results aggregated by professional and demographic dimensions. (b) Results aggregated by research fields. The faction is centered by the mean values.

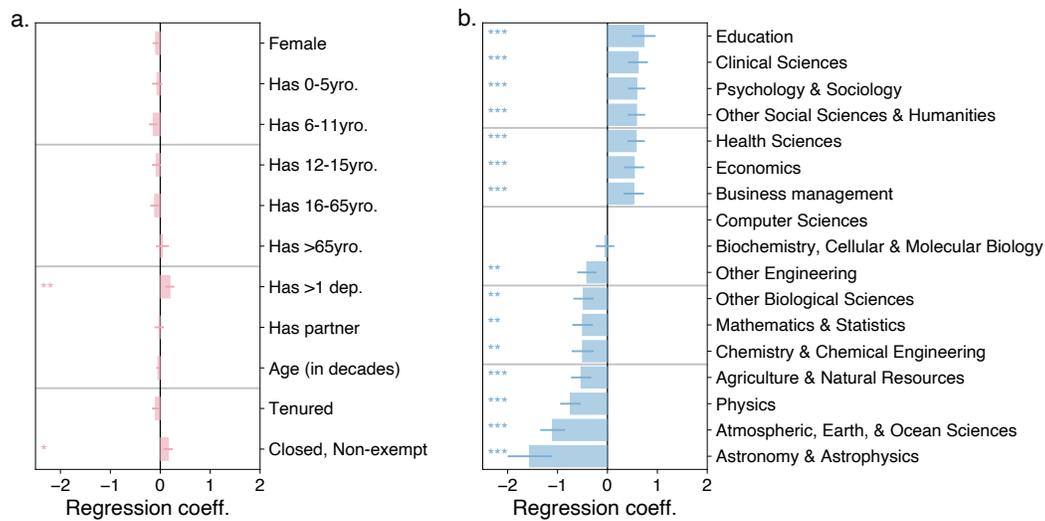

**Figure S13. Results of a probit regression that predicts whether scientists worked on COVID-related research in 2020.** (a) Regression coefficients of professional and demographic variables. (b) Regression coefficients of fields. The field "computer sciences" is used as the treatment when including the field dummy in the model. Error bars indicate standard errors, and starts indicate significant levels: *p<0.1, **p<0.05, ***p<0.01.

# SI Tables

**Table S1. Results of regressions on the number of new projects against the number of collaborators with controlling for professional and demographic variables and research fields.** The non-COVID dummy takes 1 if scientist haven't worked on COVID-related research in 2020 and 0 if otherwise. Robust standard errors in parentheses. Significant levels: *p<0.1, **p<0.05, ***p<0.01.

| | Dependent variable: Number of projects | | | | | | | | | |
|---|---|---|---|---|---|---|---|---|---|---|
| Variables | 2019 raw value | | 2019 logged value | | 2020 raw value | | | 2020 logged value | | |
| | (1) | (2) | (3) | (4) | (5) | (6) | (7) | (8) | (9) | (10) |
| Number of collaborators | 0.060*** (0.016) | 0.060*** (0.016) | 0.226*** (0.017) | 0.233*** (0.018) | 0.133*** (0.021) | 0.118*** (0.021) | 0.119*** (0.021) | 0.347*** (0.016) | 0.312*** (0.016) | 0.321*** (0.016) |
| Non Cov | | | | | | -0.878*** (0.106) | -0.850*** (0.109) | | -0.254*** (0.026) | -0.238*** (0.028) |
| Female | | -0.363*** (0.135) | | -0.069*** (0.023) | | | -0.401*** (0.098) | | | -0.104*** (0.026) |
| Has 0-5yro. | | 0.118 (0.264) | | 0.075** (0.038) | | | -0.019 (0.147) | | | -0.020 (0.040) |
| Has 6-11yro. | | 0.083 (0.170) | | -0.011 (0.032) | | | -0.063 (0.138) | | | -0.012 (0.034) |
| Has 12-15yro. | | -0.053 (0.219) | | 0.008 (0.034) | | | 0.117 (0.139) | | | 0.033 (0.035) |
| Has 16-65yro. | | 0.073 (0.165) | | 0.008 (0.033) | | | 0.239* (0.135) | | | 0.036 (0.035) |
| Has >65yro. | | 0.148 (0.236) | | 0.000 (0.050) | | | -0.129 (0.163) | | | -0.010 (0.049) |
| Has >1 dep. | | 0.128 (0.183) | | 0.013 (0.033) | | | -0.128 (0.128) | | | -0.018 (0.036) |
| Has partner | | 0.064 (0.149) | | 0.000 (0.033) | | | 0.018 (0.147) | | | 0.018 (0.036) |
| Age (in decades) | | -0.179*** (0.057) | | -0.040*** (0.012) | | | -0.113** (0.054) | | | -0.029** (0.013) |
| Tenured | | -0.118 (0.148) | | 0.001 (0.026) | | | -0.151 (0.111) | | | -0.045 (0.028) |
| Closed, Non-exempt | | 0.194 (0.271) | | -0.007 (0.034) | | | -0.060 (0.117) | | | 0.002 (0.036) |
| Constant | 2.307*** (0.077) | 2.813*** (0.404) | 0.830*** (0.024) | 0.871*** (0.086) | 1.417*** (0.056) | 2.035*** (0.103) | 2.591*** (0.390) | 0.500*** (0.018) | 0.699*** (0.027) | 0.764*** (0.097) |
| Dummy (Field) | No | Yes | No | Yes | No | No | Yes | No | No | Yes |
| Observations | 2,074 | 2,074 | 2,074 | 2,074 | 2,074 | 2,074 | 2,074 | 2,074 | 2,074 | 2,074 |
| F | 14.76 | 4.288 | 169.0 | 12.24 | 39.59 | 62.61 | 7.754 | 474.3 | 313.9 | 28.54 |
| Adjust R2 | 0.017 | 0.034 | 0.105 | 0.148 | 0.100 | 0.134 | 0.152 | 0.223 | 0.258 | 0.284 |
| RMSE | 2.572 | 2.549 | 0.502 | 0.490 | 2.109 | 2.069 | 2.047 | 0.554 | 0.542 | 0.532 |

**Table S2. Results of regressions on the absolute and percentage changes in new projects against the corresponding changes in collaborators with controlling for professional and demographic variables and research fields.** The non-COVID dummy takes 1 if scientist haven't worked on COVID-related research in 2020 and 0 if otherwise. Robust standard errors in parentheses. Significant levels: *p<0.1, **p<0.05, ***p<0.01.

| Variables | Dependent variable: Change in projects (2020 vs 2019) | | | | | | | | | |
|---|---|---|---|---|---|---|---|---|---|---|
| | Absolute change | | | | | Percentage change | | | | |
| | (1) | (2) | (3) | (4) | (5) | (6) | (7) | (8) | (9) | (10) |
| Change in collaborators | 0.107*** | 0.104*** | 0.102*** | 0.101*** | 0.102*** | 0.295*** | 0.279*** | 0.277*** | 0.277*** | 0.278*** |
| | (0.018) | (0.018) | (0.018) | (0.017) | (0.017) | (0.019) | (0.020) | (0.020) | (0.020) | (0.020) |
| Non Cov | | -0.308** | -0.331** | -0.344*** | -0.449*** | | -0.158*** | -0.163*** | -0.159*** | -0.163*** |
| | | (0.125) | (0.129) | (0.121) | (0.110) | | (0.030) | (0.030) | (0.030) | (0.033) |
| Female | | | -0.156 | -0.155 | -0.057 | | | -0.056** | -0.065** | -0.049 |
| | | | (0.099) | (0.106) | (0.123) | | | (0.027) | (0.029) | (0.030) |
| Has 0-5yro. | | | -0.376*** | -0.193 | -0.184 | | | -0.061* | -0.093** | -0.096** |
| | | | (0.121) | (0.234) | (0.233) | | | (0.036) | (0.045) | (0.046) |
| Has 6-11yro. | | | | -0.177 | -0.177 | | | | -0.006 | -0.007 |
| | | | | (0.157) | (0.157) | | | | (0.039) | (0.039) |
| Has 12-15yro. | | | | 0.161 | 0.171 | | | | 0.009 | 0.005 |
| | | | | (0.182) | (0.185) | | | | (0.039) | (0.039) |
| Has 16-65yro. | | | | 0.181 | 0.196 | | | | 0.028 | 0.030 |
| | | | | (0.142) | (0.140) | | | | (0.040) | (0.040) |
| Has >65yro. | | | | -0.174 | -0.223 | | | | 0.054 | 0.049 |
| | | | | (0.199) | (0.206) | | | | (0.055) | (0.056) |
| Has >1 dep. | | | | -0.199 | -0.207 | | | | -0.012 | -0.013 |
| | | | | (0.166) | (0.160) | | | | (0.039) | (0.040) |
| Has partner | | | | -0.041 | -0.059 | | | | 0.034 | 0.035 |
| | | | | (0.140) | (0.139) | | | | (0.037) | (0.037) |
| Age (in decades) | | | | 0.017 | 0.040 | | | | -0.023 | -0.021 |
| | | | | (0.046) | (0.048) | | | | (0.015) | (0.015) |
| Tenured | | | | -0.038 | -0.090 | | | | -0.053* | -0.065** |
| | | | | (0.141) | (0.129) | | | | (0.031) | (0.032) |
| Closed, Non-exempt | | | | -0.264 | -0.239 | | | | 0.021 | 0.026 |
| | | | | (0.247) | (0.244) | | | | (0.041) | (0.041) |
| Constant | -0.635*** | -0.436*** | -0.298** | -0.249 | -0.121 | -0.240*** | -0.138*** | -0.102*** | 0.022 | -0.009 |
| | (0.051) | (0.115) | (0.143) | (0.344) | (0.376) | (0.015) | (0.024) | (0.026) | (0.085) | (0.113) |
| Dummy (Field) | No | No | No | No | Yes | No | No | No | No | Yes |
| Observations | 1,868 | 1,868 | 1,868 | 1,868 | 1,868 | 1,868 | 1,868 | 1,868 | 1,868 | 1,868 |
| F | 35.53 | 22.77 | 13.94 | 7.812 | 4.794 | 233.0 | 142.3 | 75.47 | 24.82 | 12.28 |
| Adjust R2 | 0.0686 | 0.0720 | 0.0761 | 0.0771 | 0.0788 | 0.207 | 0.219 | 0.221 | 0.221 | 0.222 |
| RMSE | 2.255 | 2.251 | 2.246 | 2.245 | 2.242 | 0.577 | 0.573 | 0.572 | 0.572 | 0.572 |

# SI References